\documentclass[conference,a4paper]{IEEEtran}
\IEEEoverridecommandlockouts 

\usepackage{amsmath,amssymb,amsbsy,amsxtra,graphicx}
\usepackage{array,multirow}
\usepackage{balance}
\usepackage{booktabs}
\usepackage{url}
\usepackage{cite}
\usepackage{psfrag}
\usepackage{dcolumn}
\usepackage{cite}
\usepackage{bm}
\usepackage{caption}
\usepackage{graphicx}
\usepackage{verbatim}
\usepackage{colortbl}
\usepackage{algorithmic}
\usepackage{epstopdf}
\usepackage{textpos}

\graphicspath{{./Images/}{./Images/Fig1/}{./Images/Fig2/}{./Images/Fig3/}{./Images/Fig4/}{./Images/Fig5/}{./Images/Fig6/}}

\begin{document}
\title{Perceptual Compressive Sensing based on \\ Contrast Sensitivity Function: Can we avoid non-visible redundancies acquisition?}
\author{\IEEEauthorblockN{Seyed Hamid Safavi and Farah Torkamani-Azar}
\IEEEauthorblockA{Faculty of Electrical and Computer Engineering,\\
Shahid Beheshti University, Tehran, Iran, 1983969411\\ Email: {\tt \{h\_safavi, f-torkamani\}@sbu.ac.ir} }}

\maketitle

\begin{abstract}
Conventional compressive sensing (CS) attempts to acquire the most important part of a signal directly. In fact, CS avoids acquisition of existed \textit{statistical redundancies} of a signal. Since the sensitivity of the human eye is different for each frequency, in addition to statistical redundancies, there exist \textit{perceptual redundancies} in an image which human eye could not detect them. In this paper, we propose a novel CS approach in which the acquisition of non-visible information is also avoided. Hence, we could expect a better compression performance. We deploy the weighted CS idea to consider these perceptual redundancies in our model. Moreover, the block-based compressed sensing is favorable since it has some advantages: (a) It needs low memory to store the sensing matrix and sparsifying basis. (b) All blocks can be reconstructed in parallel. Therefore, we apply our proposed scheme in the block-based framework to make it practical to use. Simulation results verify the superiority of our proposed method compared to the other state-of-the-art methods. 
\end{abstract}

\IEEEpeerreviewmaketitle 

\section{Introduction}
Compression is a fundamental part of the signal processing. Since there exist a lot of redundancies when we sample a signal using Nyquist rate, therefore, it needs to be compressed. It is known that two general types of redundancies exist in an image: statistical and perceptual redundancies. Statistical redundancies could be eliminated using popular approaches like JPEG which incorporates the transform coding and variable length coding \cite{pennebaker1992jpeg}. Perceptual redundancies depend on the human visual system (HVS). The sensitivity of the human eye is different for each \textit{spatial frequency} which is defined as a number of sine or square wave gratings per visual angle \cite{netravali2013digital}. There are a lot of reasearch in this direction to provide a mathematical model of human perception. In fact, the human eye is sensitive to distortion above a threshold level. Contrast sensitivity function (CSF) is proposed to model this kind of sensitivity \cite{mannos1974effects}. It is also shown that the HVS behavior is like a band pass filter for luminance frequency. Therefore, perceptual redundancies can be removed by assigning the weights to the transform coding coefficients before quantization using CSF. In addition to CSF, there are also other effects which are image dependent like luminance masking and contrast masking \cite{watson1993dctune}.

It is well known that Compressive sensing (CS) efficiently acquire the most important part of a signal by sampling below the Nyquist rate \cite{candes,donoho}. In fact, it alleviates the existed statistical redundancies. Recently, there has also been a lot of interest to adapt the weighted compressive sensing (WCS) to enhance the sparsity \cite{chartrand2008iteratively, daubechies2010iteratively,candes2008}. Iteratively reweighted least squares minimization (IRLS) approach of \cite{chartrand2008iteratively} has been proposed to benefit from regularization. The authors in \cite{daubechies2010iteratively} also proposed another IRLS approach to enhance the sparsity. In contrast to IRLS approaches, in \cite{candes2008}, Candes \textit{et al.} have proposed to design the weights based on the $\ell_1$ norm using log-sum penalty. Since no certain guarantees have been made for algorithm's success in \cite{candes2008} and the error bound is not provided, the authors in \cite{shen2013exact}, obtained a theoretical justification of minimizing log-sum penalty. However, these methods are iterative and computationally  complex than the unweighted one. Therefore, the question would be how we could design the \textit{non-iterative} weighting matrix for CS?

There are a few works that have attempted to design the weighting coefficients using HVS. In \cite{ciocoiu2015foveated}, the authors proposed foveated compressed sensing approach which utilizes HVS. This method is used for image compression and not for image acquisition since it needs foveal information. Moreover, in \cite{yang2009perceptual}, for two-dimensional discrete cosine transform (DCT) the weighting coefficients for \textit{measurement matrix} are derived from the standard JPEG quantization table by taking the inverse of the table entries and adjusting their amplitudes to a proper range. Similarly, in \cite{Safavi2017ICASSP}, we have proposed to design the weighting coefficients for \textit{enhancing sparsity}  based on the standard JPEG quantization table for a video sequence.

Motivated by the fact that the human eye is not much sensitive to the high frequency details, we could give an importance to each frequency in signal reconstruction, such that the high frequencies have a small effect on the performance. Hence, we could expect better visuality performance from the proposed scheme compared to other conventional weighted methods. One benefit that could be expected for this approach is the weighting coefficients are independent of input images which make it non-iterative and fast. In summary, different from the mentioned weighted approaches, here we propose to benefit from contrast sensitivity function to design weighting coefficients with the aim of enhancing sparsity. 

The rest of the paper will be organized as follows. Section \ref{Background} provides a brief overview on CS and WCS. Section \ref{Proposed approach} presents the proposed method. Simulation results and discussions are given in section \ref{Experiments}. Finally, section \ref{Conclusion} concludes the paper and some future directions are
presented.

\section{Background} \label{Background}
\subsection{Compressive Sensing (CS)}
Consider a signal,  $\mathbf{x}\in {{\mathbb{R}}^{N \times 1}}$ which possesses sparse representation in a domain ${\mathbf{\Psi}} \in {\mathbb{R}^{N \times N}}$, i.e. ${\mathbf{x}} = {\mathbf{\Psi s}}$ where ${\mathbf{s}} \in {\mathbb{R}^{N}}$ and ${\left\| {\mathbf{s}} \right\|_0} = k$ (${\left\| . \right\|_0}$ stands fo the $\ell_0$ norm). According to CS theory, this signal can be
recovered uniquely from linear measurements ${\mathbf{y}} \in {\mathbb{R}^{M}}$: ${\mathbf{y}} = {\mathbf{\Phi x}} = {\mathbf{\Phi \Psi s}}$ where ${\mathbf{\Phi }} \in {\mathbb{R}^{M \times N}}$ is a full-rank sensing matrix with $\left( M < N\right)$. Gaussian measurement matrices satisfy restricted isometry property (RIP) constraint with high probability if the number of measurements was, at least, equal to $M=\mathcal{O}\left( {k\log \left( {{N \mathord{\left/ {\vphantom {n k}} \right. \kern-\nulldelimiterspace} k}} \right)} \right)$ \cite{donoho}. Therefore, the signal can be uniquely recovered by solving the following minimization problem

\begin{IEEEeqnarray}{rCl}
\label{BP}
\mathbb{P}_{\ell_0}: \,\, \underset{\mathbf{s}\in {{\mathbb{R}}^{N}}}{\mathop{\min}}\,\,{{\left\| \mathbf{s} \right\|}_{0}} \qquad \nonumber \\
\textmd{s.t.} \,\, \mathbf{y}= \underbrace {{\mathbf{\Phi \Psi }}}_{\bm{\theta }} \mathbf{s}
\end{IEEEeqnarray}
Although solving problem \eqref{BP} gives the exact solution, but unfortunately, it is NP-hard requiring an exhaustive search over all $\binom{N}{k}$ possible solutions. There are some classical sparse approximation methods which find the approximate solution like Matching Pursuit (MP) \cite{mallat1993matching} and Orthogonal MP (OMP) \cite{pati1993orthogonal}, etc. However, these greedy algorithms are fast, but they did not have worthy performance. One alternative method to avoid NP-hardness of the zero norm minimization is to replace zero norm with smallest convex $\ell_p$ norm which is $\ell_1$ norm. Hence, the problem \eqref{BP} transformed into following $\ell_1$ minimization problem:
\begin{IEEEeqnarray}{rCl}
\label{f}
\mathbb{P}_{\ell_1}: \,\, \underset{\mathbf{s}\in {{\mathbb{R}}^{N}}}{\mathop{\min}}\,\,{{\left\| \mathbf{s} \right\|}_{1}} \quad \nonumber \\
\textmd{s.t.} \,\, \mathbf{y}=\bm{\theta } \mathbf{s}
\end{IEEEeqnarray}

It is proved that this optimization problem can exactly recover the $k$-sparse signals and closely approximate the compressible signals with high probability \cite{donoho,baraniuk2007compressive}. However the computational complexity of this approach is about $\mathcal{O}\left( {{N^3}} \right)$ and could be kind of slow for large size signals. Hence there are some approaches trying to find a better function for approximating the zero norm. In \cite{mohimani}, the authors have tried to directly minimize the smoothed version of zero norm which is called SL0 and is 2 or 3 orders of magnitude faster than $\ell_1$ minimization. In SL0 approach, the exponential smoothed version of the zero norm is considered. 

\begin{figure}[t]
  \centering
  \includegraphics[width=8cm,height=6cm]{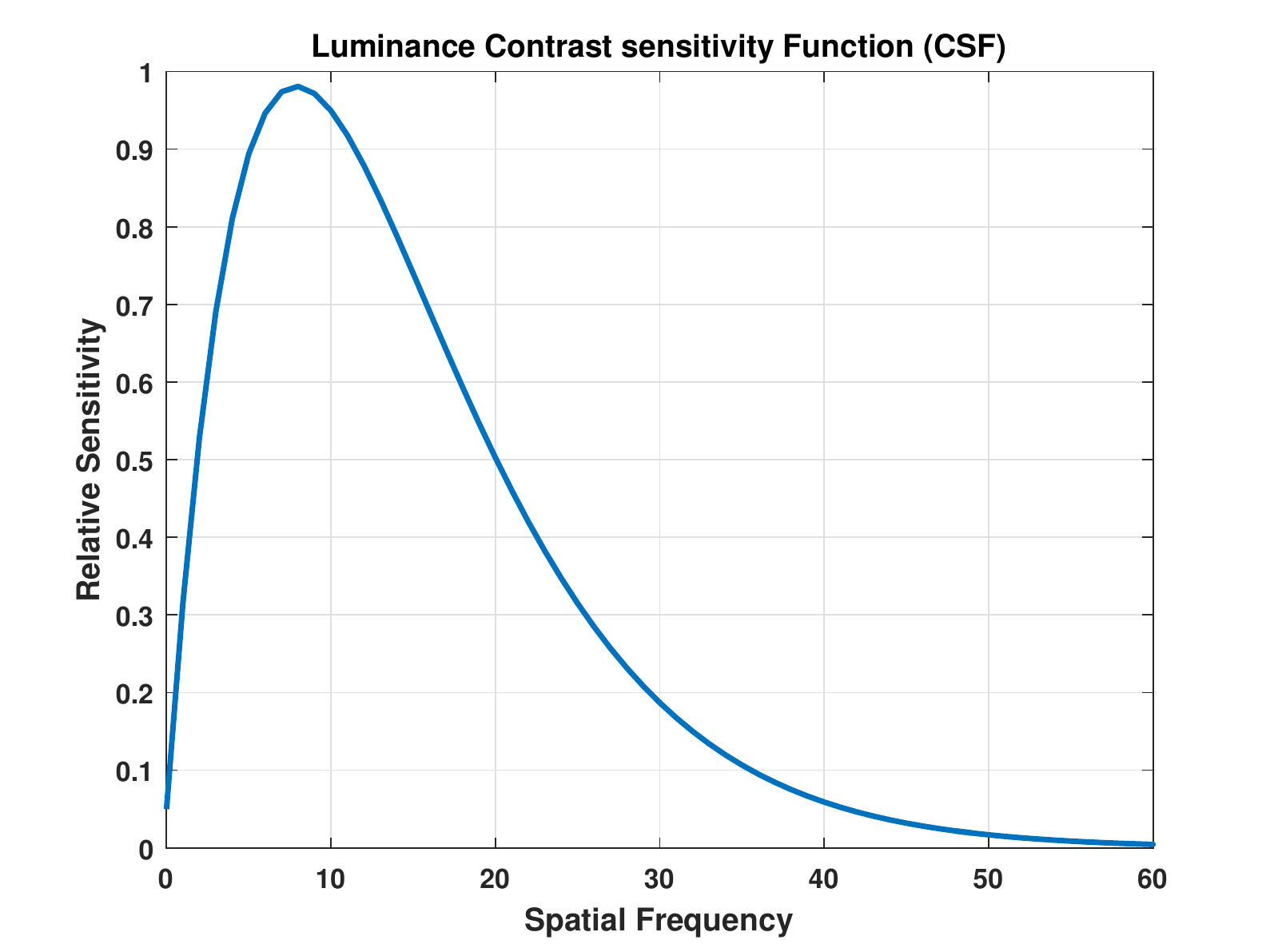}
  \caption{Contrast Sensitivity Function}
  \label{CSF_Mannos}
\end{figure}
\subsection{Weighted Compressive Sensing (WCS)}
The authors in \cite{candes2008} have tried to replace zero norm with log-sum penalty function which resulted in iterative reweighted $\ell_1$-minimization. It has shown that this approach reconstructs the exact $k$-sparse signal ($\mathbf{\Psi} = \mathbf{I}$) with a lower number of measurements than $\ell_1$-minimization algorithm, however, the computational complexity is increased because of the iterative reweighting structure. The reweighted ${\ell_1}$-minimization approach is defined as:
\begin{IEEEeqnarray}{rCl}
\label{RWL1}
\mathbb{P}_{RW\ell_1}: \,\, \underset{\mathbf{s}\in {{\mathbb{R}}^{N}}}{\mathop{\min }}\,\,\sum\limits_{i=1}^{{N}}{{{w}_{i}}\left| {{s}_{i}} \right|} \nonumber\\
\textmd{s.t.} \,\, \mathbf{y}=\bm{\theta } \mathbf{s} \,\, \quad
\end{IEEEeqnarray}
where ${{w}_{i}}>0$ denotes the weights at index $i$. To obtain the solution with the same sparsity structure of the original signal, ${w}_{i}$ should have small values on the nonzero locations of signal and significantly larger values elsewhere. Since we haven't the prior information about the location and amplitude of the non-zero elements of the sparse signal, selecting weights is done by iterative manner. Common approaches for iteratively computing weights are based on recalculating weights at every iteration using the solution of (\ref{RWL1}) at the previous iteration. Suppose ${s}_i^{\left(l\right)}$ denotes the $i$'th element of the solution of (\ref{RWL1}) for a given set of weights in $l$'th iteration. In the next iteration, the weights are updated as \cite{candes2008}:
\begin{equation}\label{RWL1_weight}
{{w}_{i}^{\left(l+1\right)}}=\frac{1}{\left| {{s}_i^{\left(l\right)}} \right|+\varepsilon }
\end{equation}
where $i=1,2,...,{N}$, and $\varepsilon $ is a positive parameter in order to provide stability especially when ${{s}_i^{\left(l\right)}}$ is zero valued. It should be noted that the iterative reweighted $\ell_1$-minimization algorithm has better performance for exactly $k$-sparse signals. However, the improvement of this approach decreases for compressible signals.

\section{Proposed approach} \label{Proposed approach}
Since the RWL1 approach \cite{candes2008} is iterative and it has a higher computational complexity compared to unweighted CS, in this research, we try to answer this question: how we could design the \textit{non-iterative} and \textit{image independent} weighting matrix of CS? Motivated by the human visual system (HVS) which state that the sensitivity of human eye is not same for different frequencies, we could give an importance to each frequency in signal reconstruction. Hence, we could expect better visuality performance from the proposed scheme compared to other state-of-the-art WCS methods. One benefit that could be expected for this approach is that the weighting coefficients are image independent which makes it non-iterative and fast.

To design a perceptual WCS scheme, first, we should know how humans see the world. There are a lot of research in this direction to provide a mathematical model of human perception. The human eye is sensitive to the \textit{spatial frequency} which is defined as a number of sine or square wave gratings per visual angle. The function that model this effect is called contrast sensitivity function (CSF). The following model for CSF is originally proposed by Mannos and Sakrison \cite{mannos1974effects} which is plotted in Fig. \ref{CSF_Mannos}.
\begin{equation} \label{Eq-CSF}
H\left( {{f_{i,j}}} \right) = 2.6\left( {0.0192 + 0.114{f_{i,j}}} \right){e^{ - {{\left( {0.114{f_{i,j}}} \right)}^{1.1}}}}
\end{equation}
where ${f_{i,j}}$ stands for the spatial frequency which is defined in the DCT domain as follows: 
\begin{equation} \label{Eq:Spatial_f}
{f_{i,j}} = \frac{1}{{2N}}\sqrt {{{\left( {\frac{i}{{{\theta _x}}}} \right)}^2} + {{\left( {\frac{j}{{{\theta _y}}}} \right)}^2}}  \quad \left( {{\text{cycles/degree}}} \right)
\end{equation}
where $i,j = 0, \cdots, N-1$. Here, $\theta _x$, $\theta _y$ denote the visual angles in the $x$ and $y$ direction and are defined as follows:

\begin{figure}[t]
\begin{minipage}[b]{.9\linewidth}
  \centering
  \centerline{\includegraphics[width=8cm,height=5.5cm]{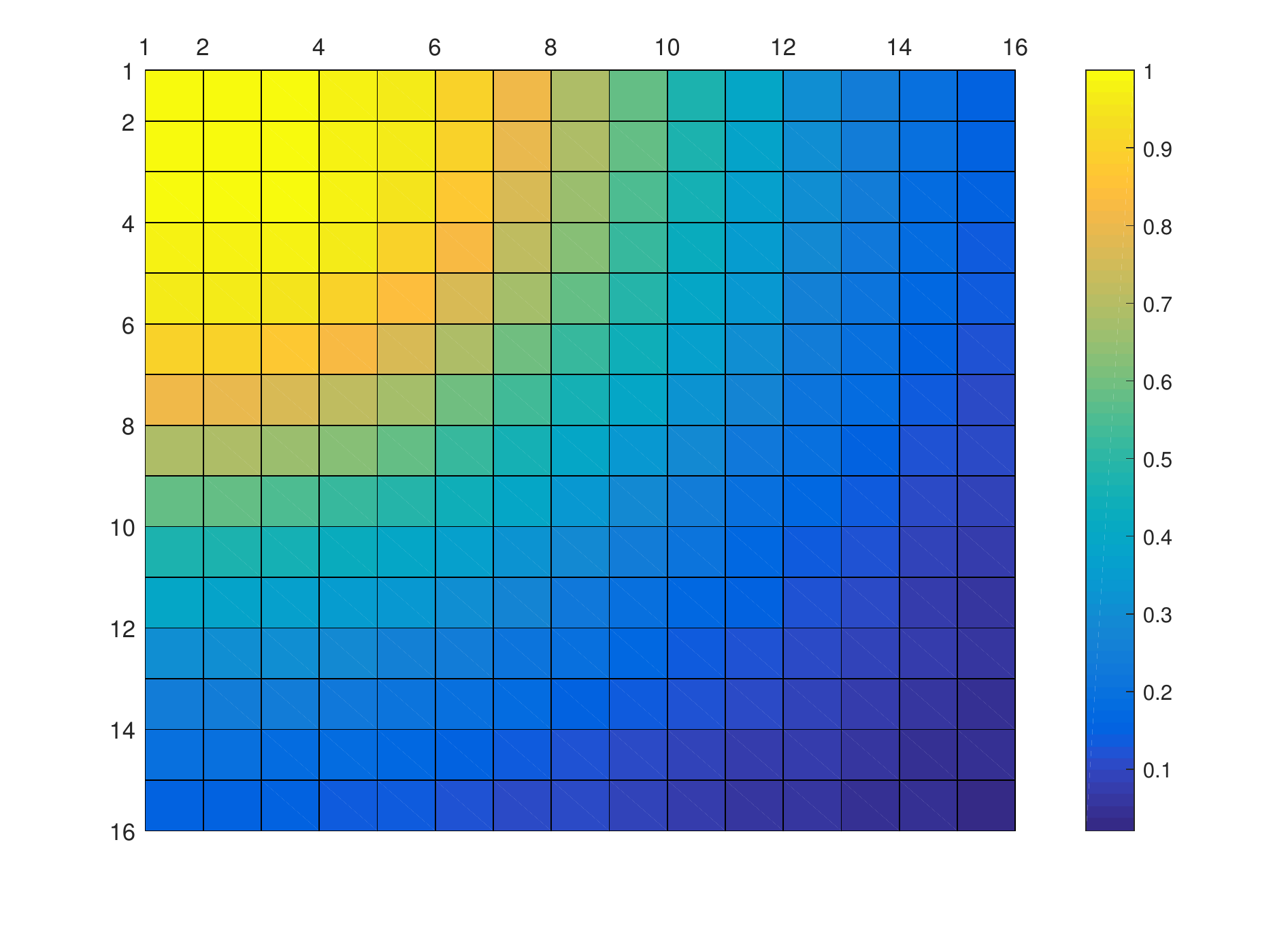}}
  \centerline{(a) Proposed scheme}\medskip
\end{minipage}
\begin{minipage}[b]{0.9\linewidth}
  \centering
  \centerline{\includegraphics[width=8cm,height=5.5cm]{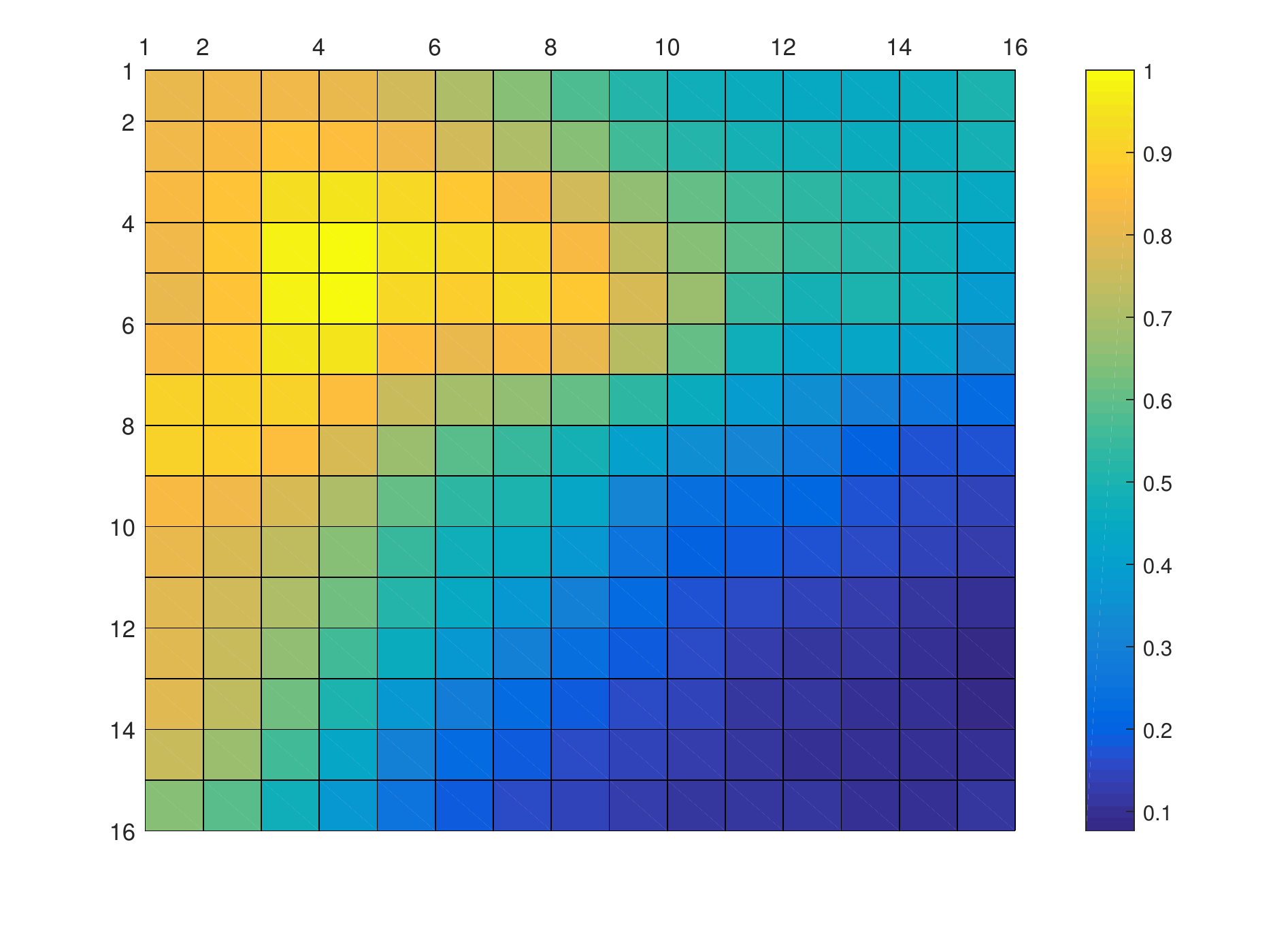}}
  \centerline{(b) The approach based on the JPEG quantization matrix \cite{Safavi2017ICASSP}}\medskip
\end{minipage}
\caption{Distribution of the CSF matrix ($\mathbf{H}$) and modified JPEG quantization matrix for $16 \times 16$ block in 2D-DCT domain: (a) Proposed scheme using CSF matrix ($\mathbf{H}$) (b) The approach based on the modified JPEG quantization matrix \cite{Safavi2017ICASSP}}
\label{Weight_Dist}
\end{figure}

\begin{equation} \label{Eq:Visual_angel}
{\theta _x} = 2\arctan \left( {\frac{{{\Lambda _x}}}{{2D}}} \right),\,\,{\theta _y} = 2\arctan \left( {\frac{{{\Lambda _y}}}{{2D}}} \right)
\end{equation}
where $D$ is the viewing distance and $\Lambda_x$, $\Lambda_y$ denote the display width/length of a pixel on the monitor. Let us define $R_{vd}$ as the ratio of viewing distance to picture height. According to the international standard ITU-R BT.500-11
\cite{ITU} (Methodology for the subjective assessment of the quality of television pictures), this ratio normally is between 3 to 6 depending on the picture size. It is also known that for most of the displays, pixel aspect ratio (PAR) is equal to 1. Therefore, we can simplify \eqref{Eq:Visual_angel} as follows:
\begin{equation}
{\theta _x} = {\theta _y} = 2\arctan \left( {\frac{1}{{2{R_{vd}}Pic_h}}} \right)
\end{equation}
where $Pic_h$ is stand for picture height. Now, let us explain how we assign the weights perceptually. It is evident that Eq. \eqref{RWL1} can be written as follows:

\begin{figure}[t]
\begin{minipage}[b]{.3\linewidth}
  \centering
  \centerline{\includegraphics[width=\linewidth]{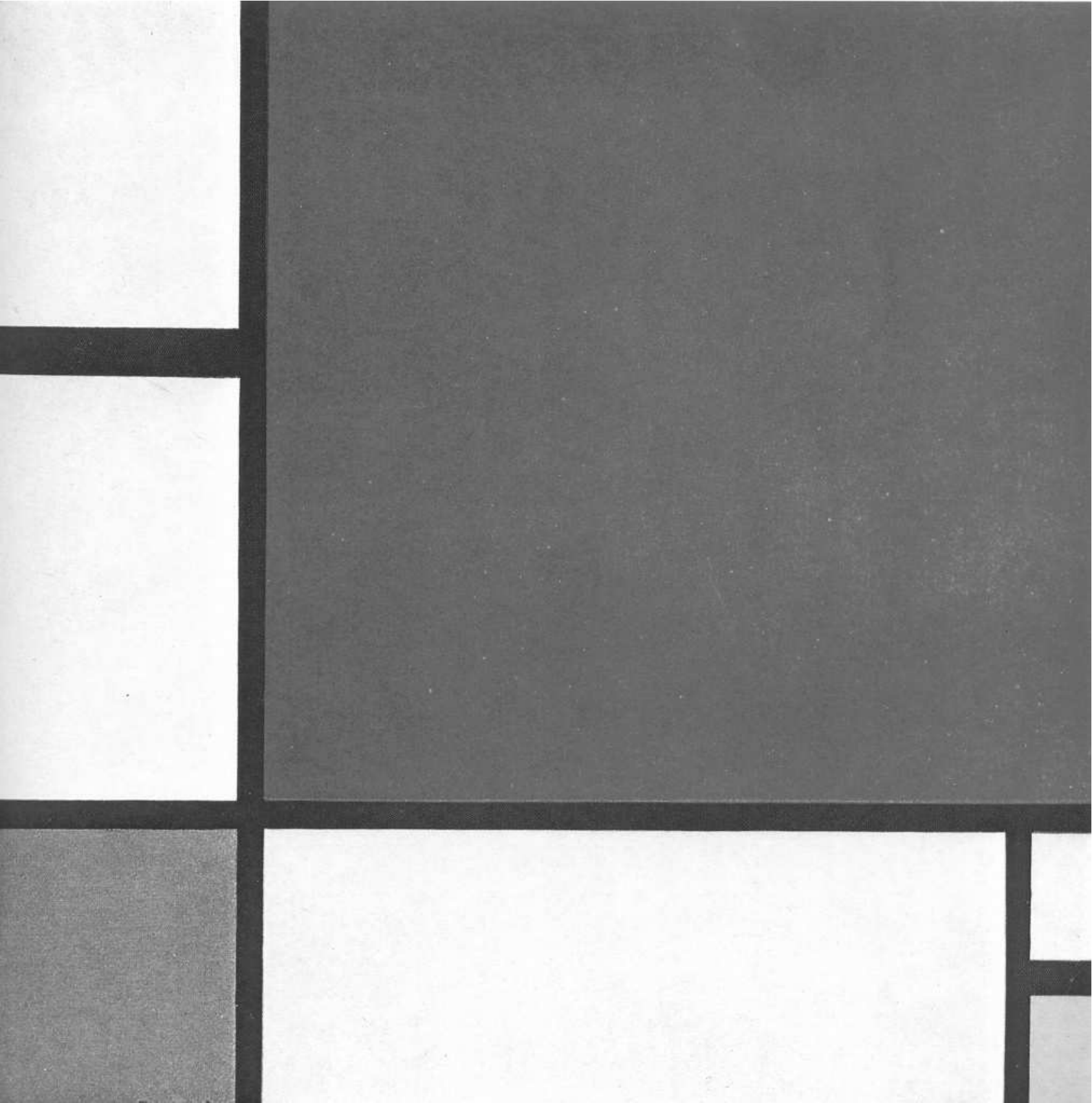}}
  \centerline{(a) Mondrian}
\end{minipage}
\begin{minipage}[b]{0.3\linewidth}
  \centering
  \centerline{\includegraphics[width=\linewidth]{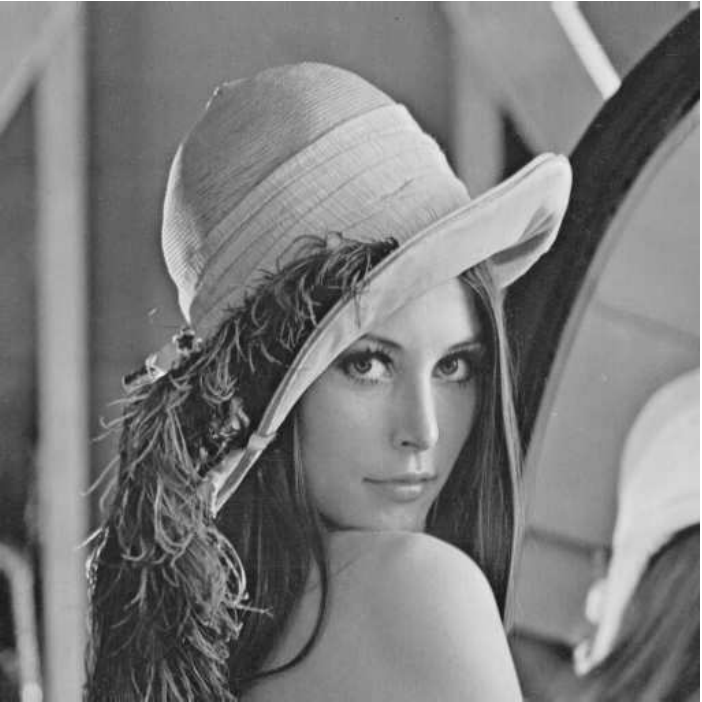}}
  \centerline{(b) Lenna}
\end{minipage}
\begin{minipage}[b]{0.3\linewidth}
  \centering
  \centerline{\includegraphics[width=\linewidth]{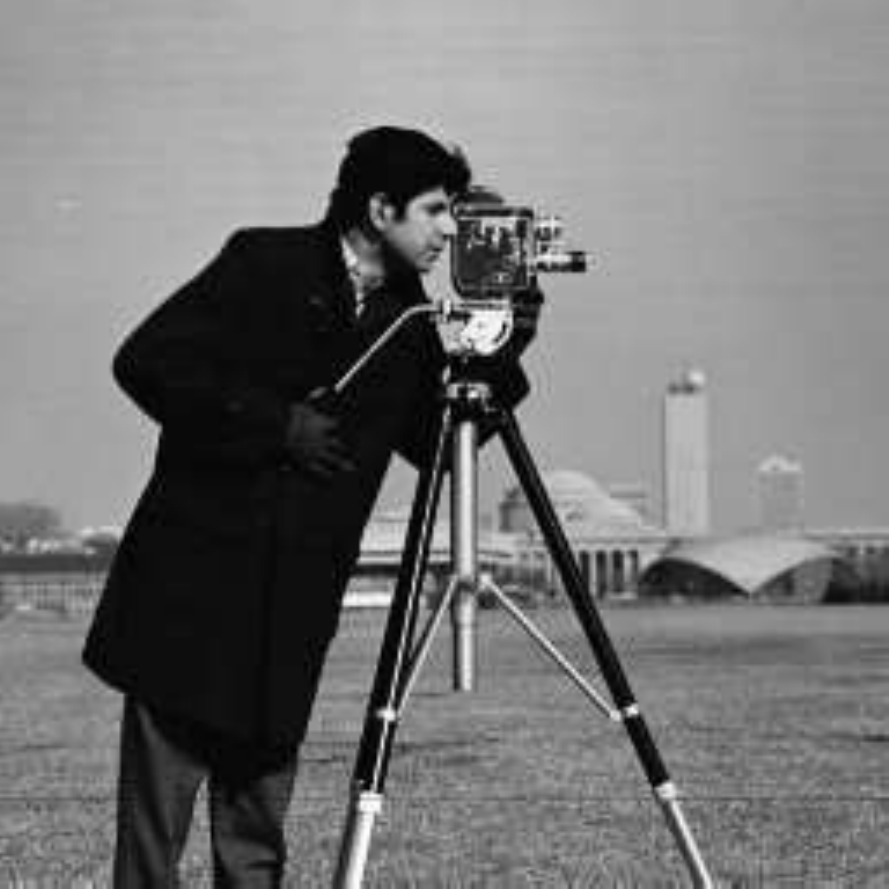}}
  \centerline{(c) Cameraman}
\end{minipage}
\begin{minipage}[b]{0.32\linewidth}
  \centering
  \centerline{\includegraphics[width=\linewidth]{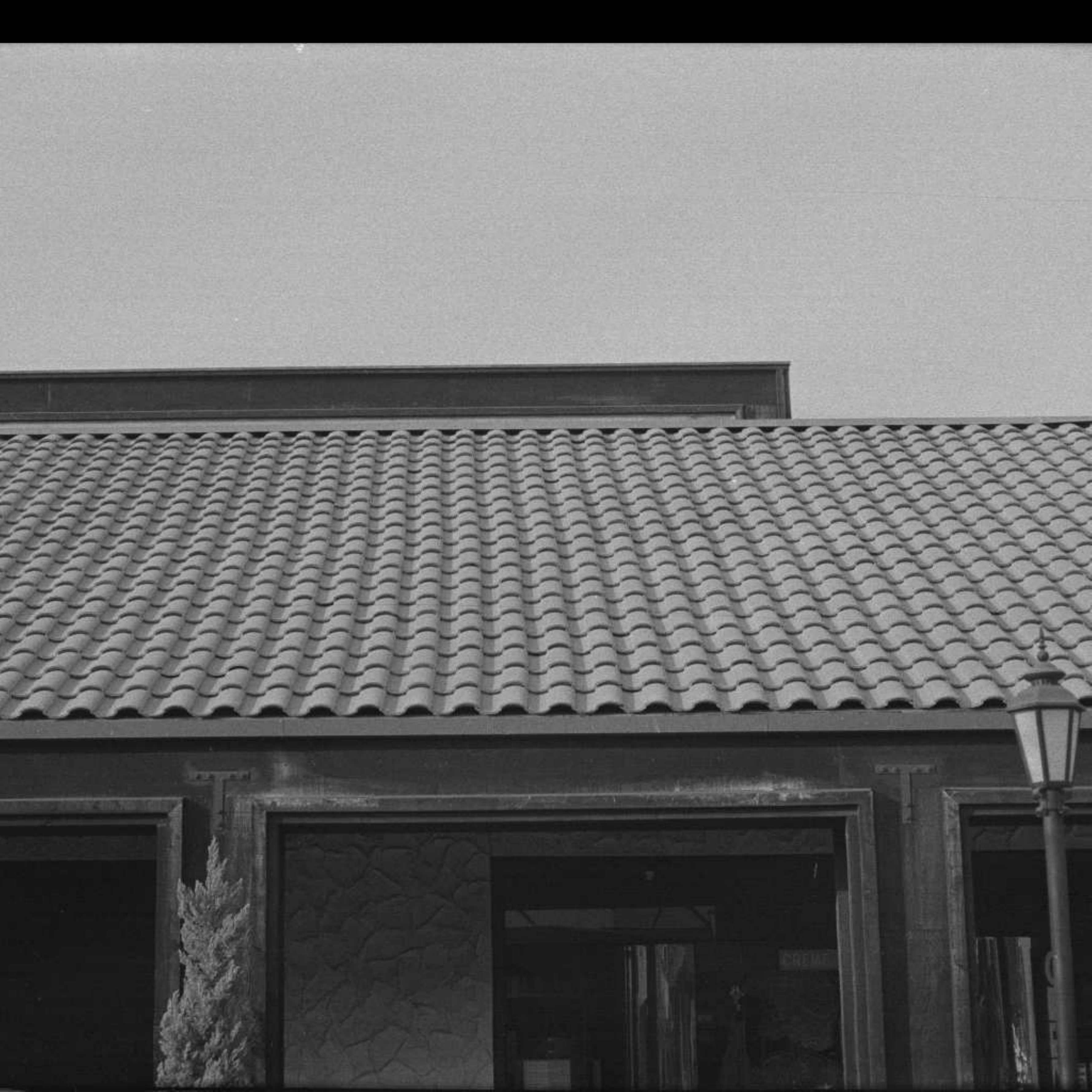}}
  \centerline{(d) Tile Roof}
\end{minipage} \hspace{0.01cm}
\begin{minipage}[b]{0.32\linewidth}
  \centering
  \centerline{\includegraphics[width=\linewidth]{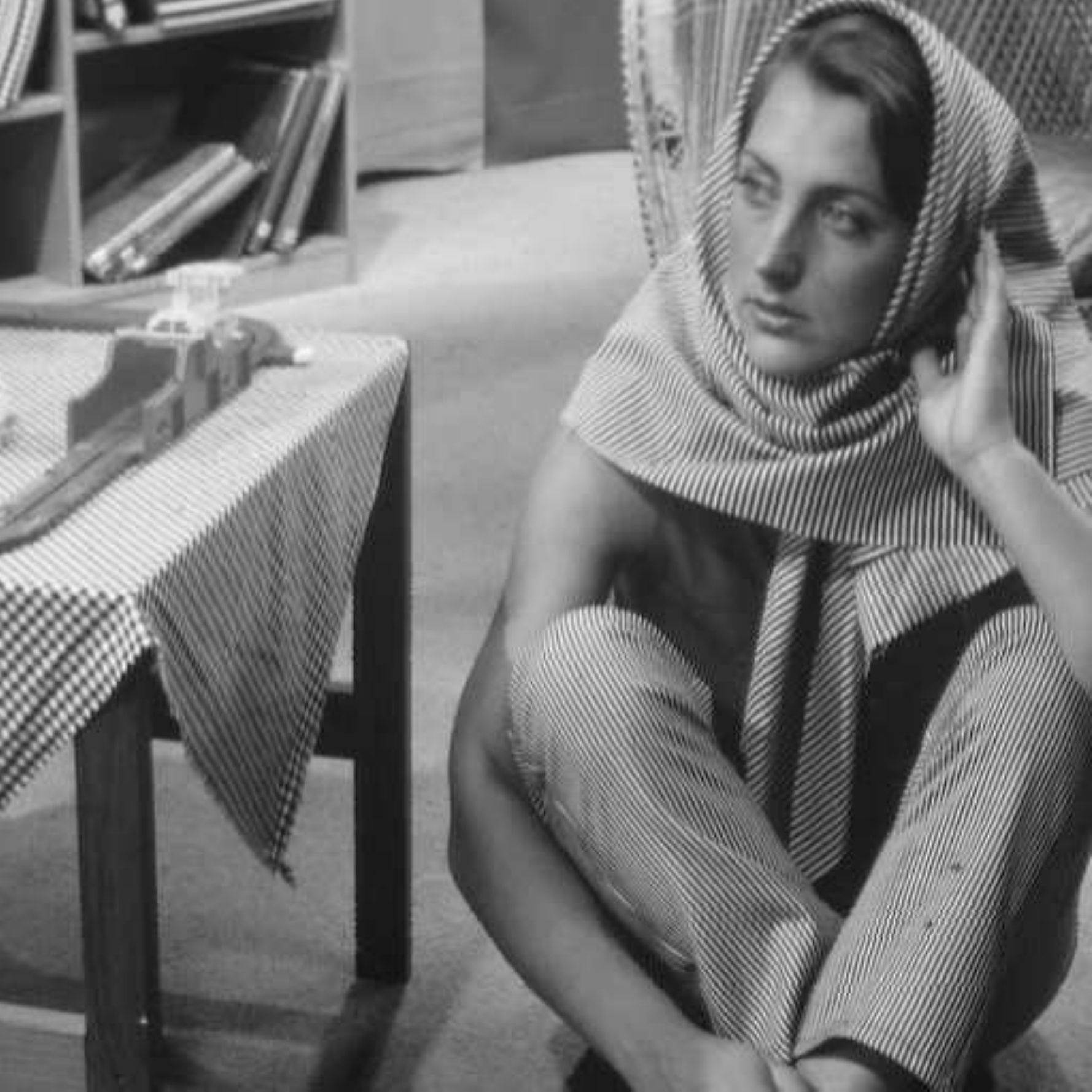}}
  \centerline{(e) Barbara}
\end{minipage} \hspace{0.01cm}
\begin{minipage}[b]{0.32\linewidth}
  \centering
  \centerline{\includegraphics[width=\linewidth]{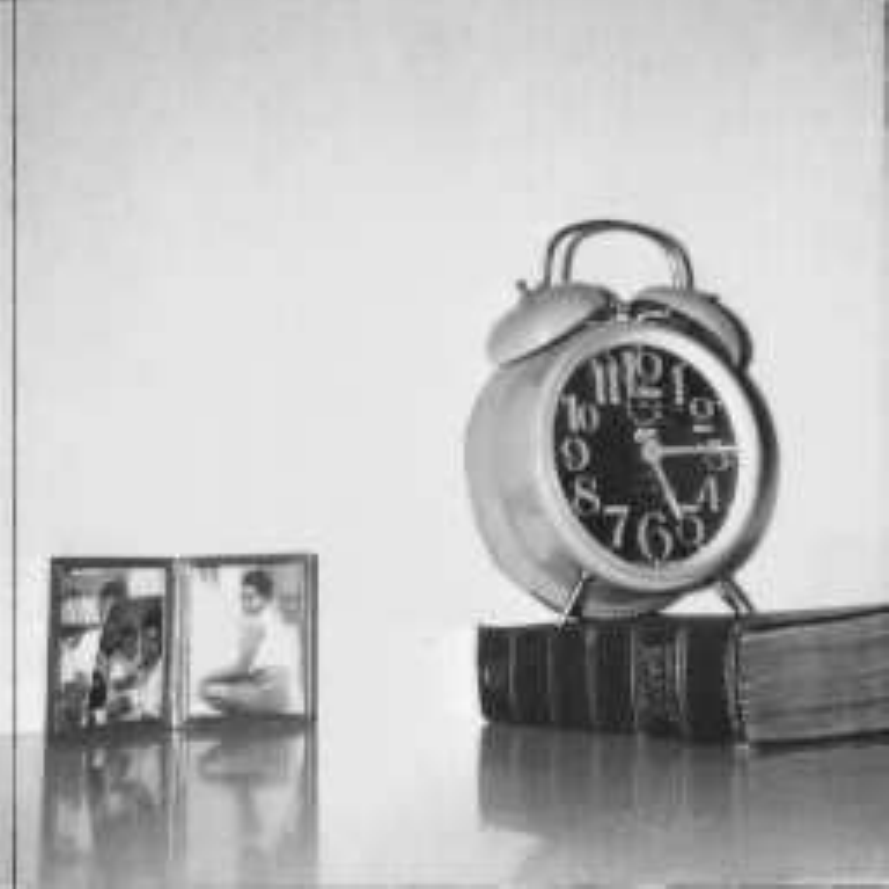}}
  \centerline{(f) Clock}
\end{minipage}
\caption{Different test images are used in the simulation results: (a) Mondrian ($512 \times 512$) (b) Lenna ($512 \times 512$) (c) Cameraman ($256 \times 256$) (d) Tile Roof ($1024 \times 1024$) (e) Barbara ($512 \times 512$) (f) Clock ($256 \times 256$).}
\label{fig3}
\end{figure}
\begin{IEEEeqnarray}{rCl}
\label{RWL1_mat}
\mathbb{P}_{RW\ell_1}: \,\, \underset{\mathbf{s}\in {{\mathbb{R}}^{N}}}{\mathop{\min }}\,\, {\left\| {{\mathbf{Ws}}} \right\|_1} \nonumber\\
\textmd{s.t.} \,\, \mathbf{y}=\bm{\theta } \mathbf{s} \,\, \quad
\end{IEEEeqnarray}
where $\mathbf{W}$ refers to a diagonal matrix with $\left\{ {w_{i}} \right\}_{i = 1}^{N}$ as a diagonal elements. Assume that $\mathbf{z} = \mathbf{Ws}$. Therefore, \eqref{RWL1_mat} is equivalent to
\begin{IEEEeqnarray}{rCl}
\label{RWL1_mat_reform}
\mathbb{P}_{RW\ell_1}: \,\, \underset{\mathbf{z}\in {{\mathbb{R}}^{N}}}{\mathop{\min }}\,\, {\left\| {{\mathbf{z}}} \right\|_1} \nonumber\\
\textmd{s.t.} \,\, \mathbf{y}=\bm{\theta } \mathbf{W}^{-1} \mathbf{z} \,\, \quad
\end{IEEEeqnarray}
Note that it is proved in \cite{candes2008} that the weighting coefficients are inversely proportional to the value of the signal, since the larger values of the signal should be less penalized. Considering this issue, we propose to use \eqref{Eq-CSF} as an inverse of weight coefficient for each element in the DCT domain. Mathematically,
\begin{equation}\label{Proposed_weight}
\mathbf{W}^{-1} = \text{Diag} \left(  \text{vec} \left(  \mathbf{H}  \right) \right)
\end{equation}

\begin{figure*}[t]
\begin{minipage}[b]{.32\linewidth}
  \centering
  \centerline{\includegraphics[width=6.4cm,height=5.2cm]{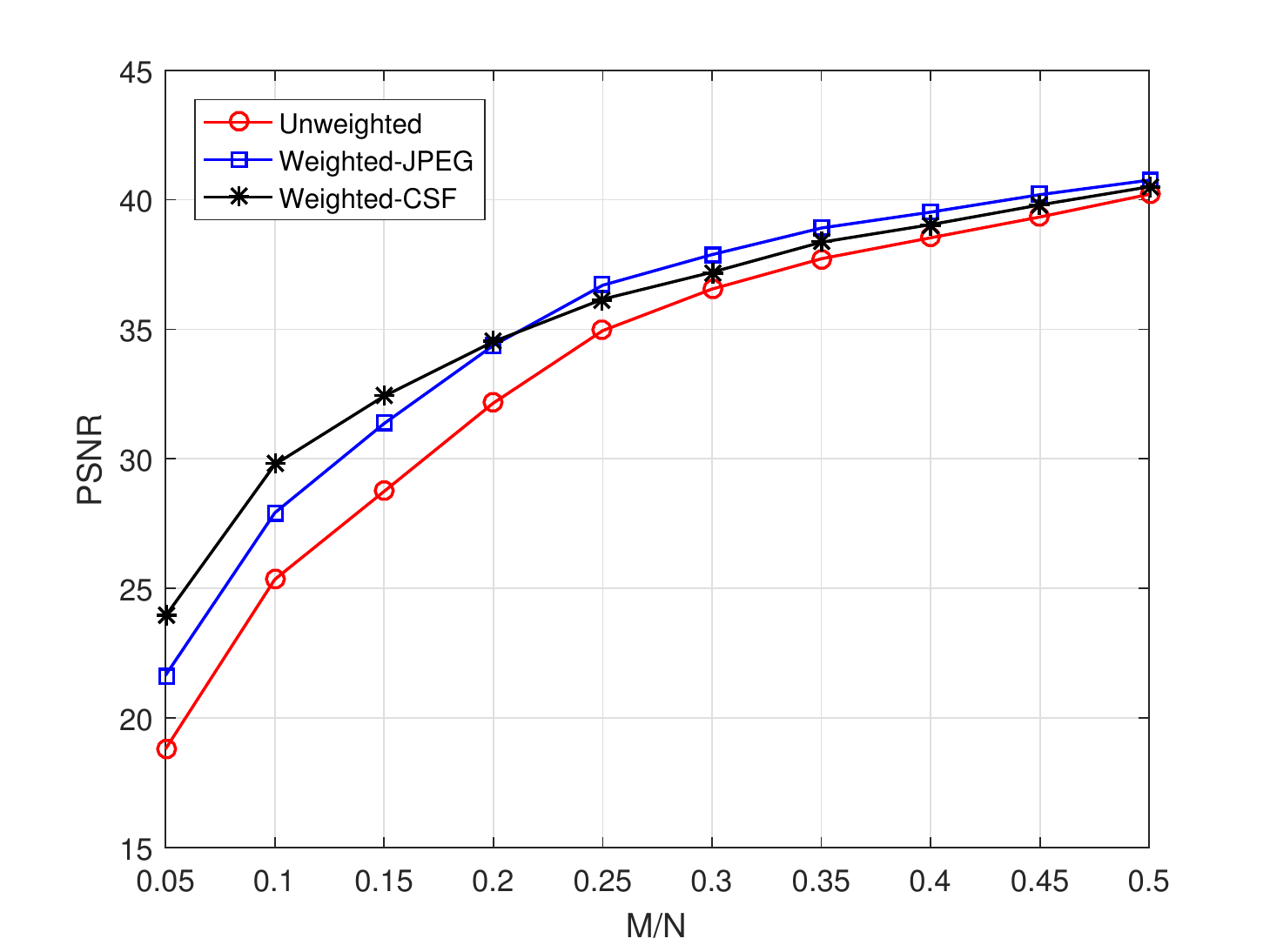}}
  \centerline{(a) Mondrian}
\end{minipage}
\begin{minipage}[b]{0.32\linewidth}
  \centering
  \centerline{\includegraphics[width=6.4cm,height=5.2cm]{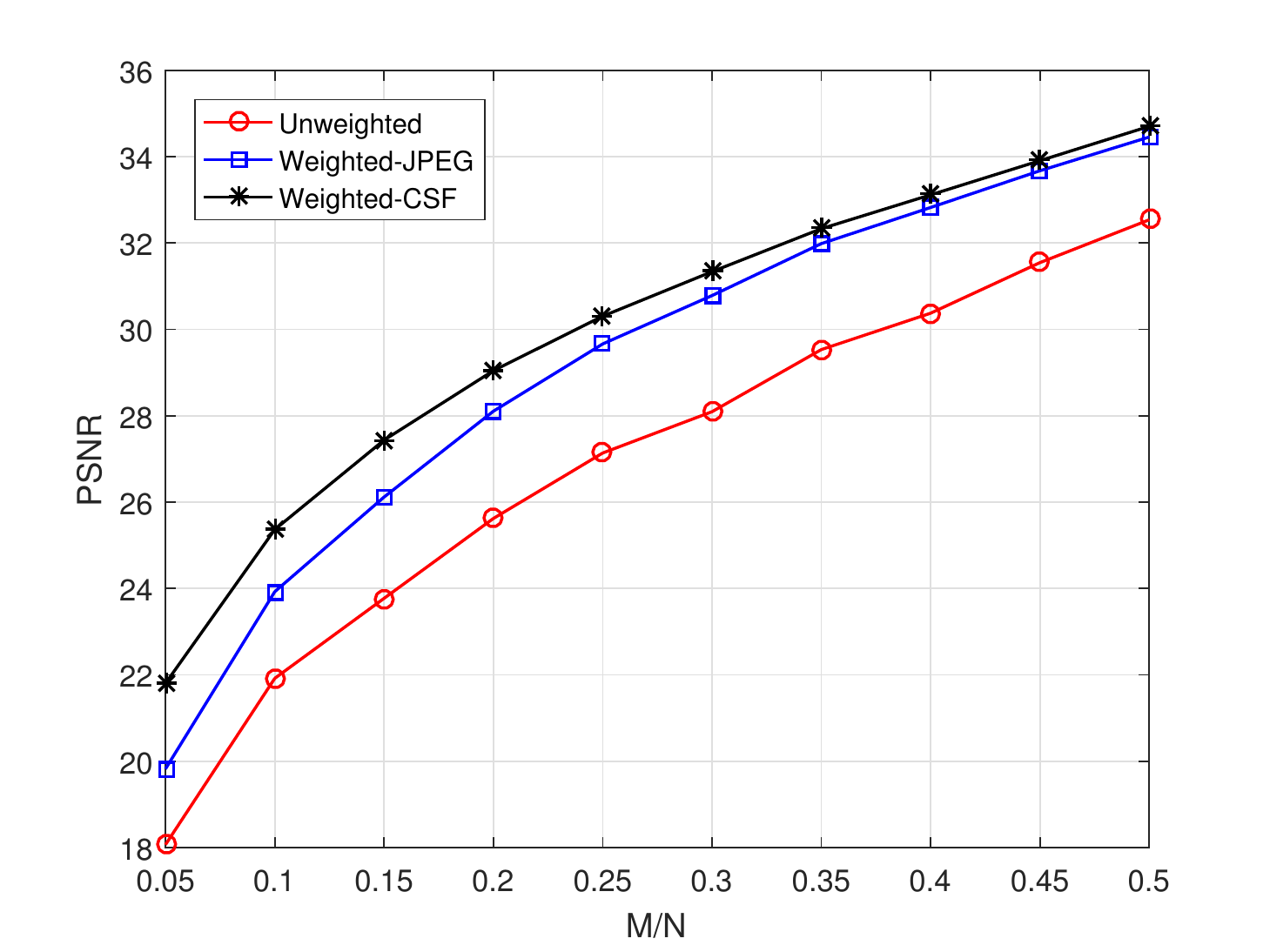}}
  \centerline{(b) Lenna}
\end{minipage}
\begin{minipage}[b]{0.32\linewidth}
  \centering
  \centerline{\includegraphics[width=6.4cm,height=5.2cm]{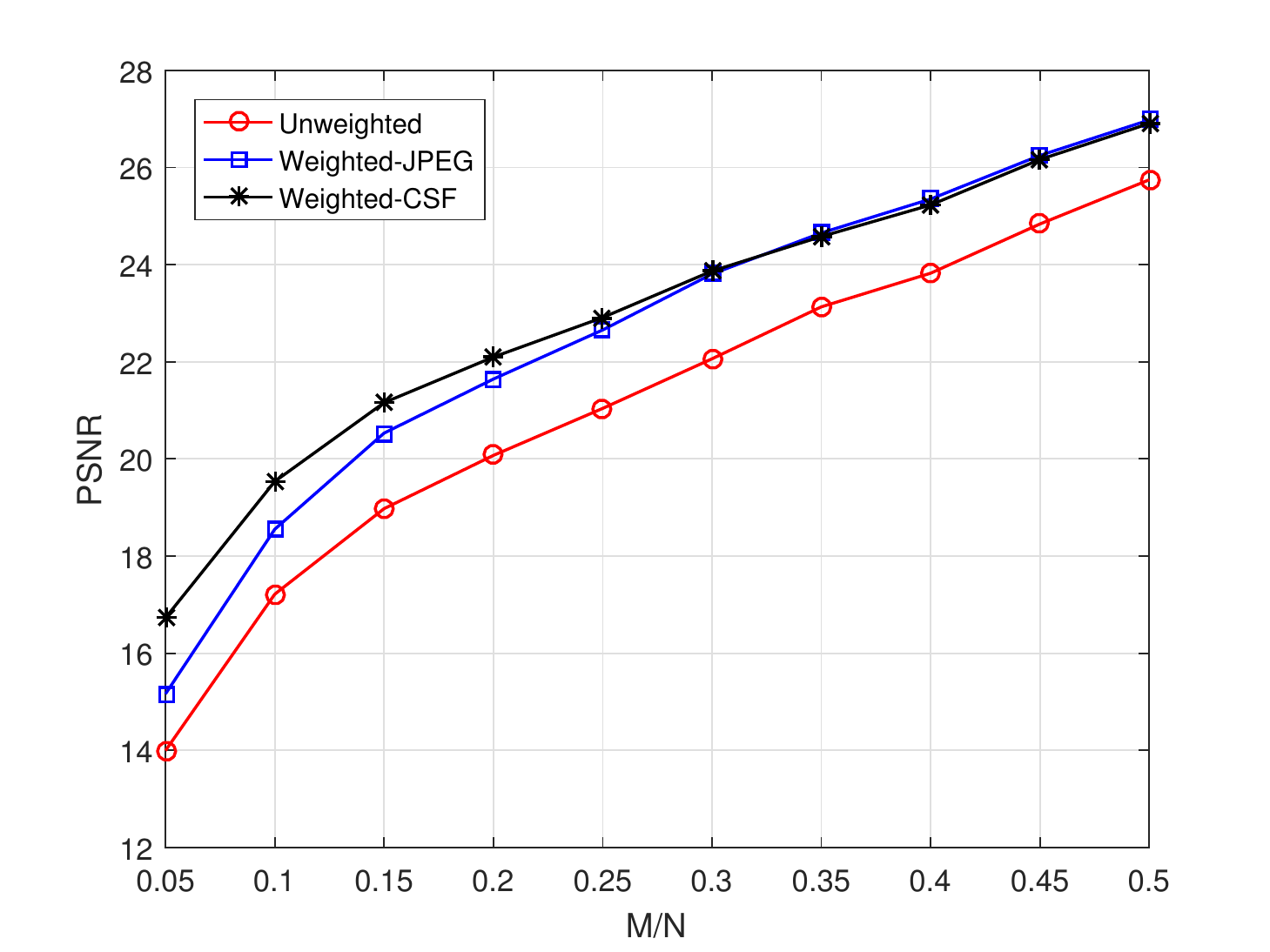}}
  \centerline{(c) Cameraman}
\end{minipage}
\begin{minipage}[b]{.32\linewidth}
  \centering
  \centerline{\includegraphics[width=6.4cm,height=5.2cm]{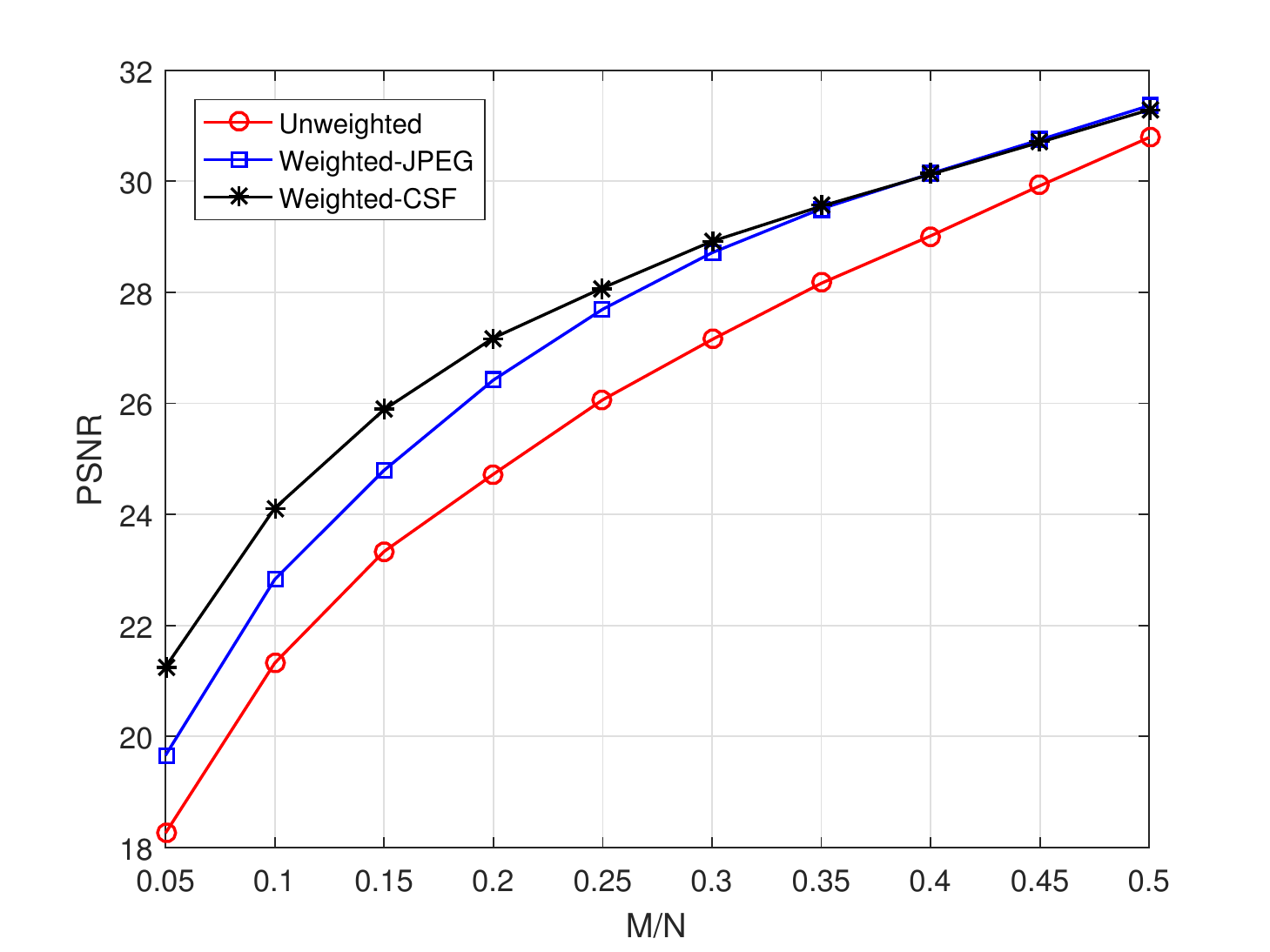}}
  \centerline{(d) Tile roof}
\end{minipage} \hspace{0.15cm}
\begin{minipage}[b]{0.32\linewidth}
  \centering
  \centerline{\includegraphics[width=6.4cm,height=5.2cm]{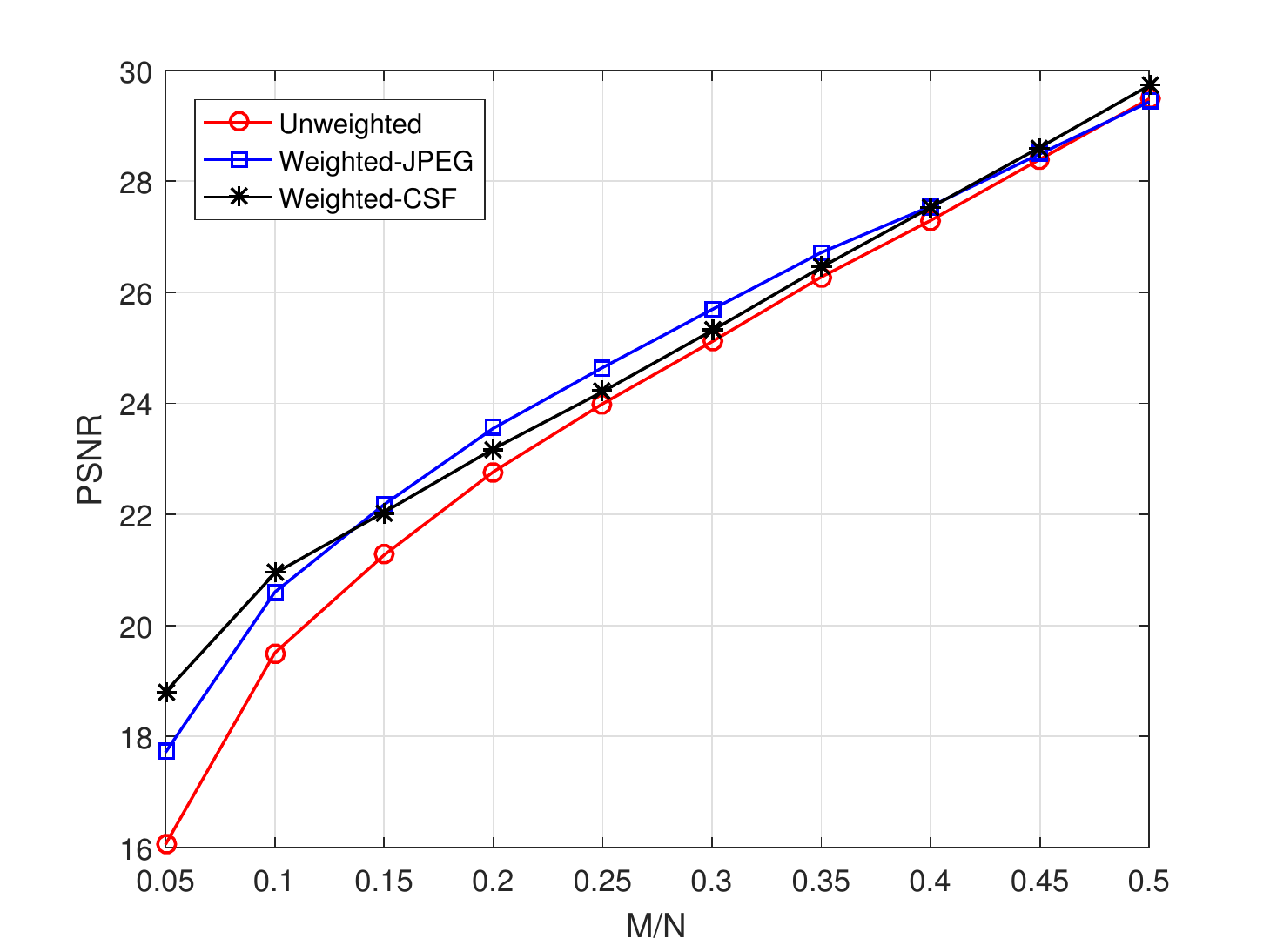}}
  \centerline{(e) Barbara}
\end{minipage}\hspace{0.15cm}
\begin{minipage}[b]{0.32\linewidth}
  \centering
  \centerline{\includegraphics[width=6.4cm,height=5.2cm]{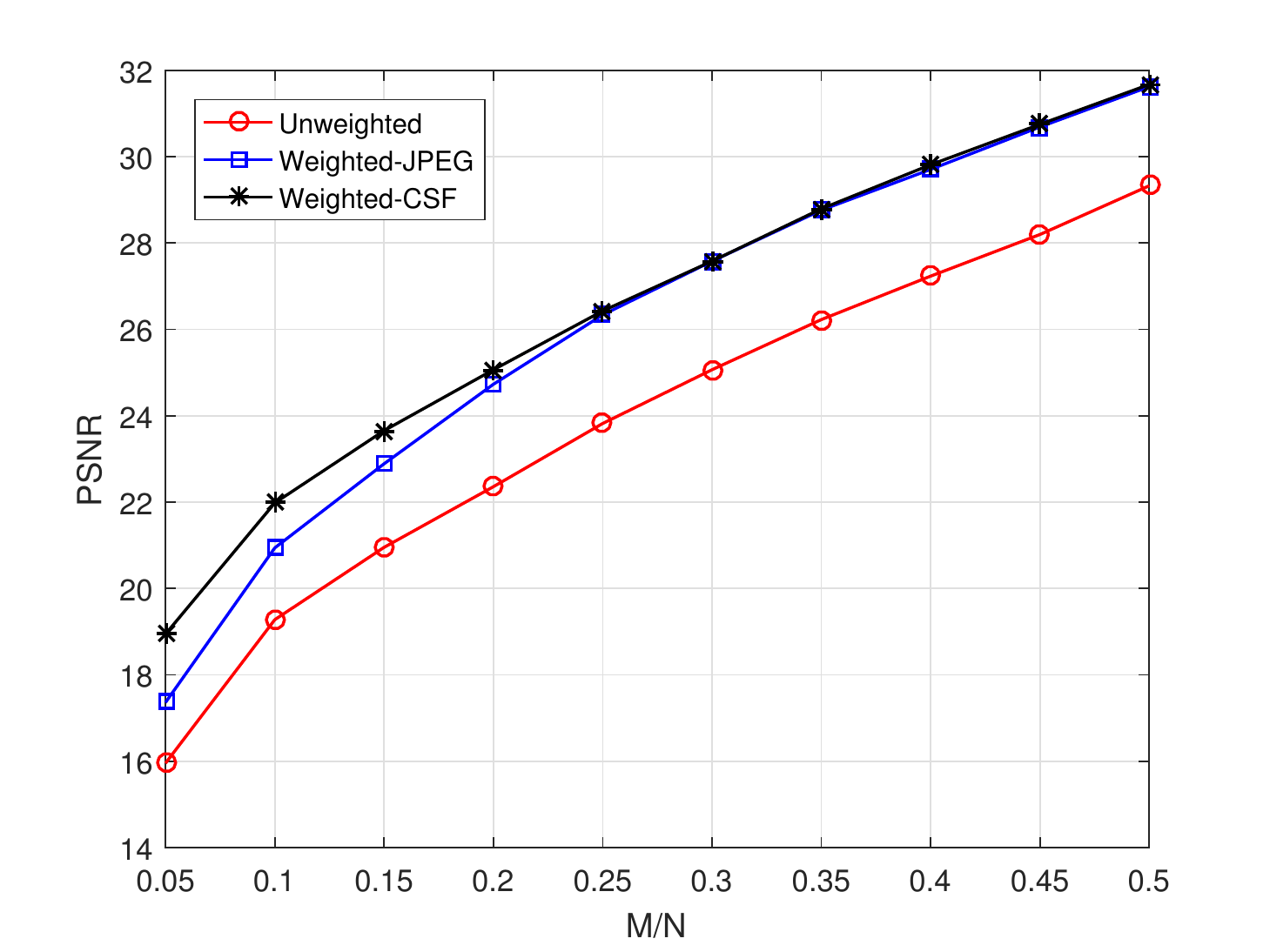}}
  \centerline{(f) Clock}
\end{minipage}
\caption{PSNR versus normalized measurements for different test images: (a) Mondrian, (b) Lenna, (c) Cameraman (d) Tile Roof (e) Barbara (f) Clock}
\label{fig4:PSNR}
\end{figure*}
\noindent where the $\left( i,j \right)$'th entry of the matrix $\mathbf{H}$ is equal to $H \left( f_{i,j} \right)$. As a last step of our proposed method, we assign the maximum value for the inverse of weight to the zero frequency, i.e. DC component since it has the highest value in the DCT domain. To better understand the way of weight allocation to each frequency in the DCT domain, in Fig. \ref{Weight_Dist}, we have visualized the distribution of the CSF matrix ($\mathbf{H}$) in our proposed scheme and we have also compared it with the one in \cite{Safavi2017ICASSP} which is based on the modified JPEG quantization matrix. As it can be seen, the CSF matrix is symmetric while the modified JPEG quantization matrix in \cite{Safavi2017ICASSP} is not symmetric. 

\section{Experiments} \label{Experiments}
In this section, we provide some simulations to validate the effectiveness of our proposed algorithm. Fig. \ref{fig3} shows six popular test images which are used in this paper, i.e. the Mondrian, Lenna, Cameraman, Tile Roof, Barbara and Clock test images with different sizes. Since the Block-based CS is used to make our approach practical, the two-dimensional image is divided into $B \times B$ blocks and then each block is sampled with an ordinary random Gaussian matrix. Also, the 2D-DCT sparsifying basis is considered in our simulations. We employ blocks of size $B = 16$ for our simulations. Also, the SparseLab software \cite{donoho2007sparselab} is used for solving the $\ell_1$-minimization problems. Finally, we have measured the accuracy of the reconstructed signal using the peak signal-to-noise ratio (PSNR) and structural similarity (SSIM) \cite{SSIM}.

\begin{figure*}[t]
\begin{minipage}[b]{.32\linewidth}
  \centering
  \centerline{\includegraphics[width=6.4cm,height=5.2cm]{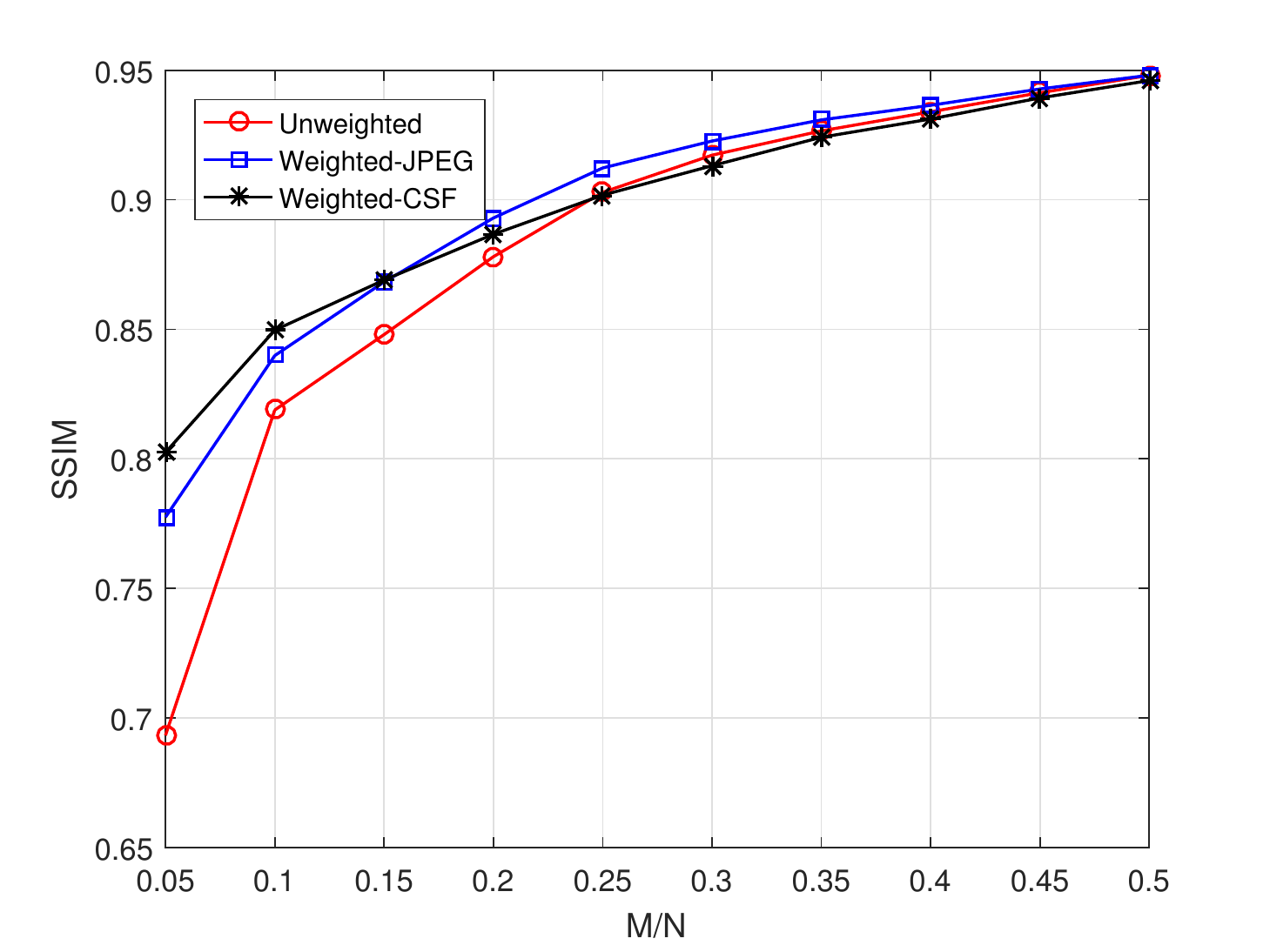}}
  \centerline{(a) Mondrian}
\end{minipage}
\begin{minipage}[b]{0.32\linewidth}
  \centering
  \centerline{\includegraphics[width=6.4cm,height=5.2cm]{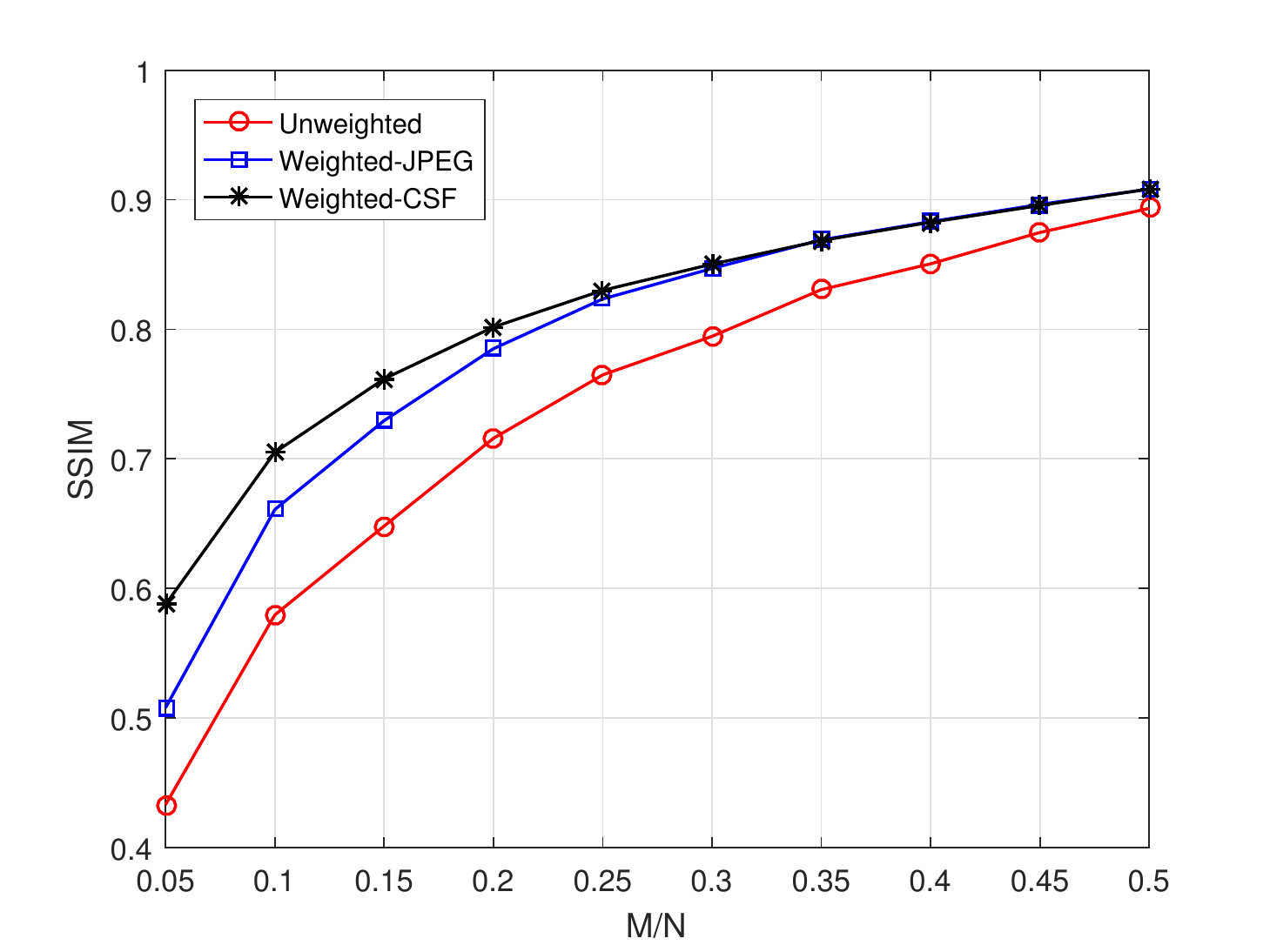}}
  \centerline{(b) Lenna}
\end{minipage}
\begin{minipage}[b]{0.32\linewidth}
  \centering
  \centerline{\includegraphics[width=6.4cm,height=5.2cm]{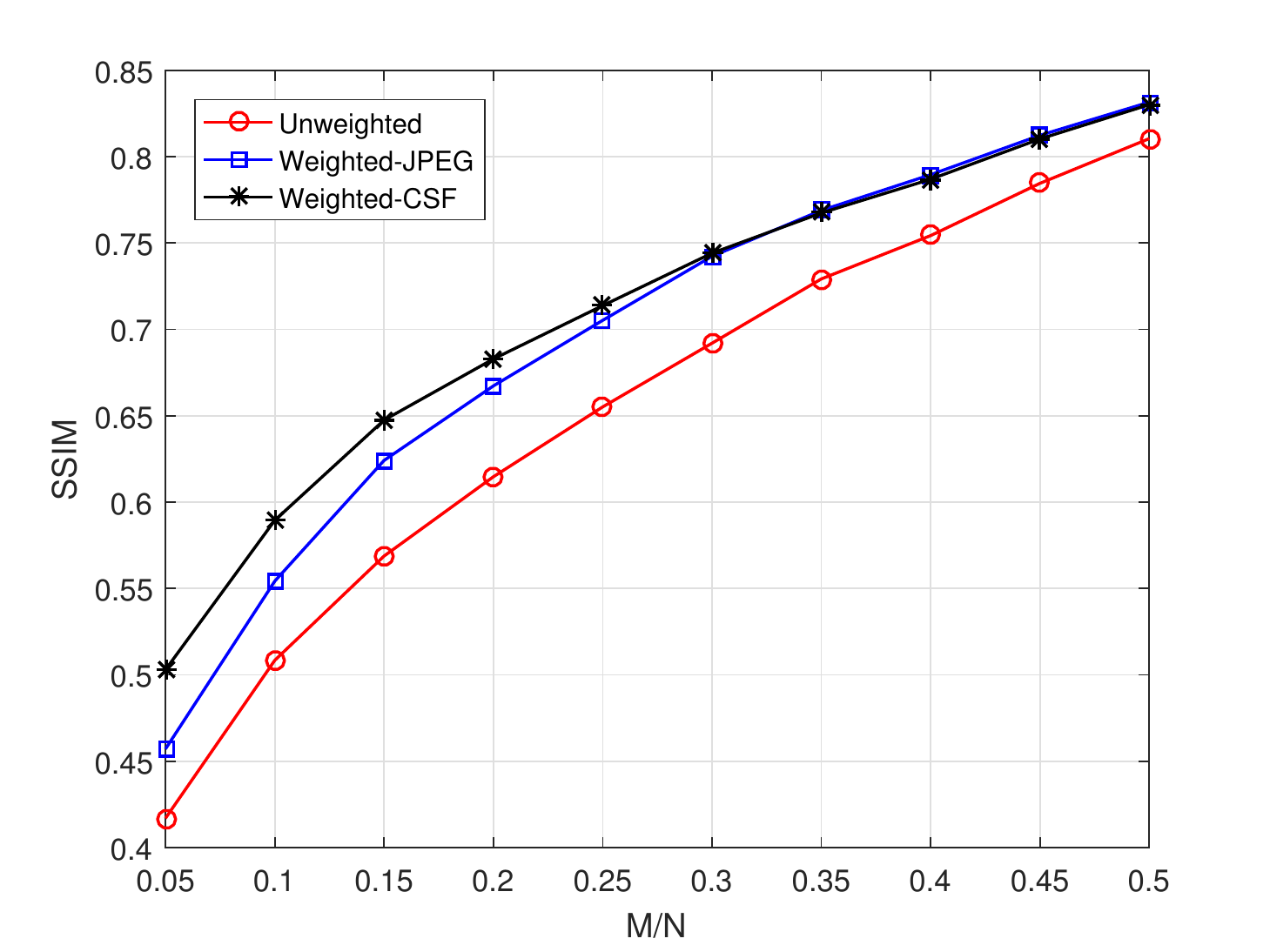}}
  \centerline{(c) Cameraman}
\end{minipage}
\begin{minipage}[b]{.32\linewidth}
  \centering
  \centerline{\includegraphics[width=6.4cm,height=5.2cm]{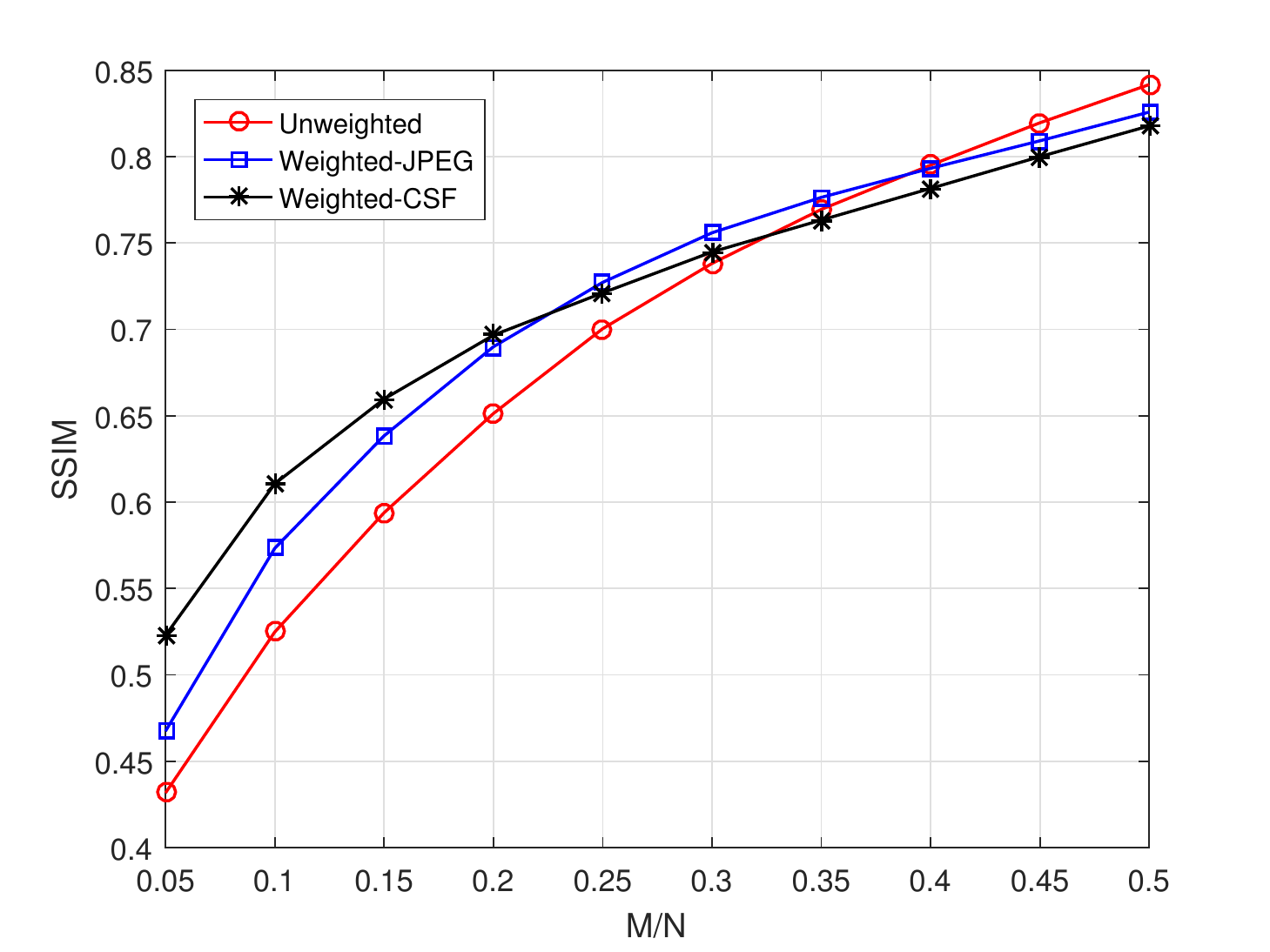}}
  \centerline{(d) Tile roof}
\end{minipage} \hspace{0.15cm}
\begin{minipage}[b]{0.32\linewidth}
  \centering
  \centerline{\includegraphics[width=6.4cm,height=5.2cm]{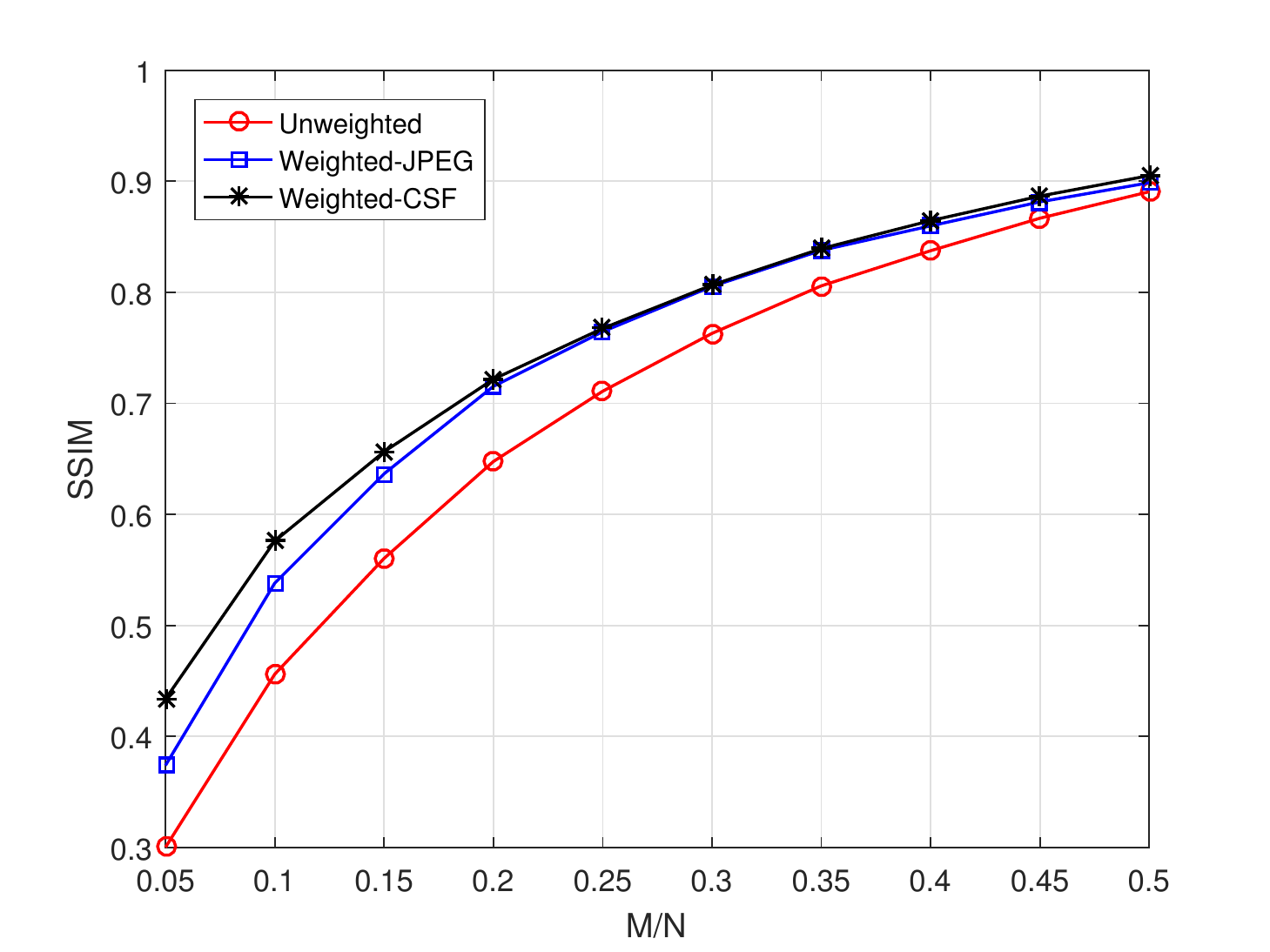}}
  \centerline{(e) Barbara}
\end{minipage}\hspace{0.15cm}
\begin{minipage}[b]{0.32\linewidth}
  \centering
  \centerline{\includegraphics[width=6.4cm,height=5.2cm]{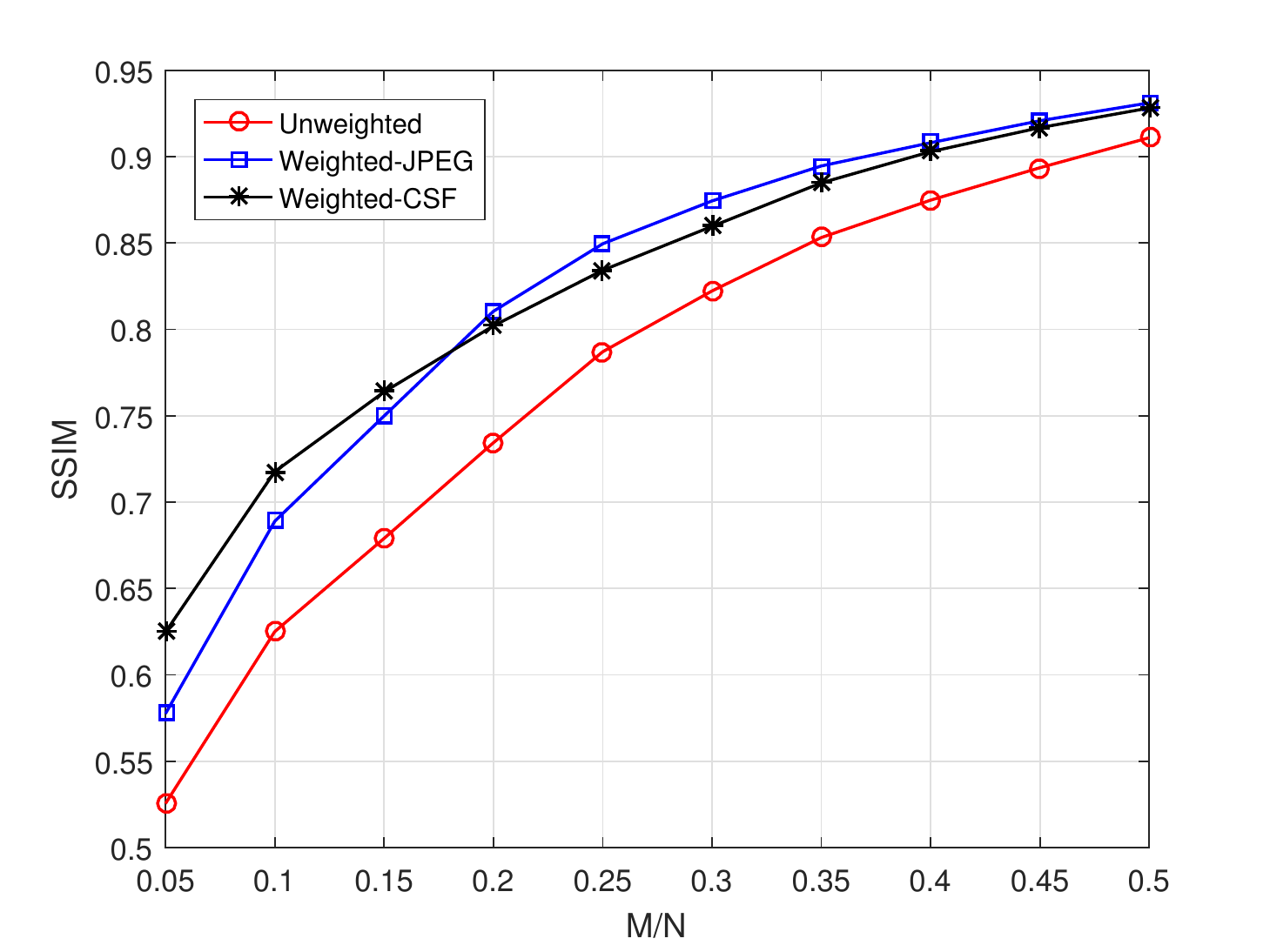}}
  \centerline{(f) Clock}
\end{minipage}
\caption{SSIM versus normalized measurements for different test images: (a) Mondrian, (b) Lenna, (c) Cameraman (d) Tile Roof (e) Barbara (f) Clock}
    \label{fig5:SSIM}
\end{figure*}

\begin{figure*}[t]
\begin{minipage}[b]{.45\linewidth}
  \centering
  \centerline{\includegraphics[width=6cm,height=6cm]{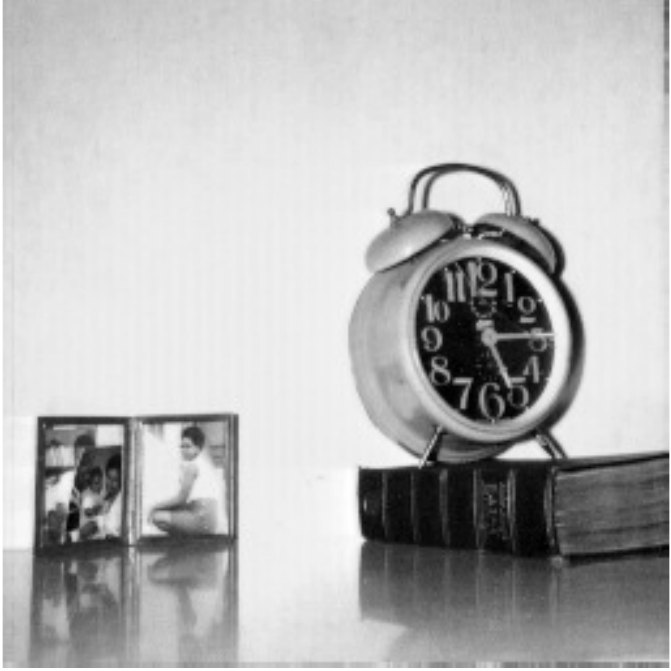}}
  \centerline{(a) Original}
\end{minipage}
\begin{minipage}[b]{0.45\linewidth}
  \centering
  \centerline{\includegraphics[width=6cm,height=6cm]{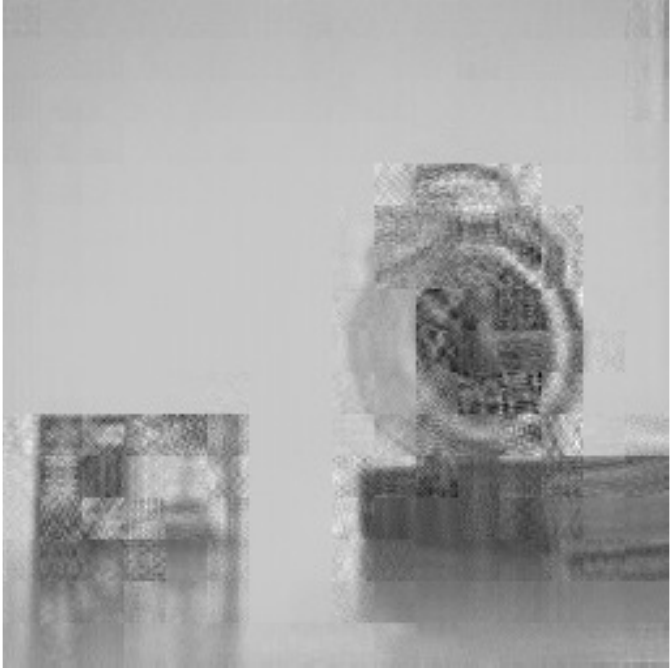}}
  \centerline{(b) Unweighted}
\end{minipage}
\begin{minipage}[b]{0.45\linewidth}
  \centering
  \centerline{\includegraphics[width=6cm,height=6cm]{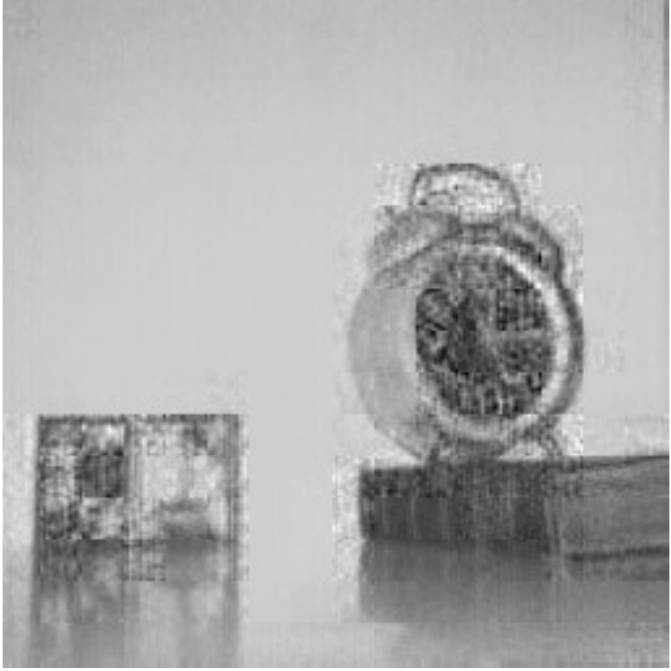}}
  \centerline{(c) Weighted based on JPEG quantization matrix}
\end{minipage} \hspace{1.55cm}
\begin{minipage}[b]{.45\linewidth}
  \centering
  \centerline{\includegraphics[width=6cm,height=6cm]{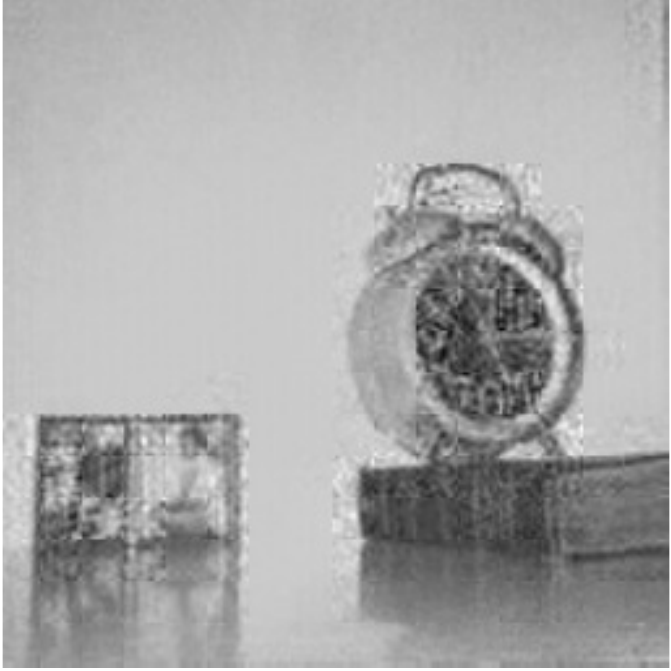}}
  \centerline{(d) Proposed weighted based on CSF}
\end{minipage}
    \caption{Comparison of visual results of the Clock test image reconstruction with  $M = 0.2N$ for different algorithms: (a) Original, (b) Unweighted, (c) Weighted based on JPEG quantization matrix (d) Proposed weighted based on CSF}
    \label{fig6:visuality}
\end{figure*}

Fig. \ref{fig4:PSNR} and \ref{fig5:SSIM} compare the PSNR and SSIM of our proposed scheme with the conventional CS and the approach of \cite{Safavi2017ICASSP} respectively. Here, we do not include the results of RWL1 \cite{candes2008} and IRLS \cite{chartrand2008iteratively, daubechies2010iteratively} since we have already shown in \cite{Safavi2017ICASSP} that their performance is not competitive with the unweighted approach for compressible signals. From Fig. \ref{fig4:PSNR} and \ref{fig5:SSIM}, it can be seen that our method is competitive with \cite{Safavi2017ICASSP} which is based on the JPEG standard quantization matrix. We also observed that when the available number of measurements is low, then our approach has the best performance among the other methods. Comparing the test images, we observe that the reconstruction performance is greatly improved for images that have more perceptual redundancies like the Lenna, Cameraman, and Clock test images. 
Fig. \ref{fig6:visuality} verified that the proposed CSF-based perceptual CS algorithm has better visual result in comparison to other methods. These results are plotted with $M = 0.2N$. Our approach has better compressibility due to avoiding acquisition of perceptual redundancies. Therefore when the number of measurements is fixed, the performance of our method becomes better compared to other methods. It should be noted that we have also observed that by increasing the block size, the performance of all schemes are improved since the blocking artifact is reduced.

\section{Conclusion} \label{Conclusion}
In this paper, we propose a novel weighted compressive sensing approach based on contrast sensitivity function. The aim of conventional CS approaches is to avoid acquisition of statistical redundancies existed in the signal. Here, we design some perceptual weights to avoid acquisition of non-visible redundancies. In fact, these weights are used to enhance the sparsity. Moreover, in contrast to other weighted CS approaches, our scheme is non-iterative and fast. We also present some simulations to verify the effectiveness of our proposed scheme compared to other state-of-the-art methods. We use contrast sensitivity function to design weights since the aim of this paper is to design non-iterative weights. As a future work, we can design perceptual weights based on CSF in the wavelet domain. We can also benefit from luminance adaption and contrast masking to design iterative approaches which avoid non-visible redundancies and depends on the image content. 


\begin{textblock*}{3cm}(4.1cm,-23cm)
	\makebox[0.1\columnwidth]{$25^{\rm th}$ Iranian Conference on
		Electrical Engineering (ICEE2017)}
\end{textblock*}
\end{document}